\documentclass[a4paper,11pt]{article}
\usepackage{pos}
\usepackage[utf8]{inputenc}
\usepackage{graphicx}
\usepackage{float}
\usepackage{amsmath}  
\usepackage{mathtools}
\usepackage{dsfont}
\usepackage{array}
\usepackage{bm,fixmath}
\usepackage{subfigure}
\usepackage{mathrsfs}
\usepackage{pifont}
\usepackage{multirow}
\usepackage{upgreek}
\usepackage{xcolor}
\usepackage{bm}
\usepackage{bbm}
\usepackage{physics}
\usepackage{comment}
\usepackage{appendix}
\DeclarePairedDelimiter\set\{\}
\usepackage{verbatim}
\usepackage{slashed}
\usepackage{cleveref}

\title{3+1D \texorpdfstring{\boldmath{$\theta$}-}{}Term on the Lattice from the Hamiltonian Perspective}

\author*[a]{Angus Kan}

\author[b,c]{Lena Funcke}

\author[d]{Stefan Kühn}

\author[a]{Luca Dellantonio}

\author[a]{\mbox{Jinglei Zhang}}

\author[a,e]{\mbox{Jan F. Haase}}

\author[a,c]{Christine A. Muschik}

\author[f]{Karl Jansen}

\affiliation[a]{Institute for Quantum Computing and Department of Physics \& Astronomy, University of Waterloo, Waterloo, Ontario, N2L 3G1, Canada}

\affiliation[b]{Center for Theoretical Physics, Co-Design Center for Quantum Advantage, and NSF AI Institute for Artificial Intelligence and Fundamental Interactions, Massachusetts Institute of Technology, 77 Massachusetts Avenue, Cambridge, MA 02139, USA}

\affiliation[c]{Perimeter Institute for Theoretical Physics, 31 Caroline Street North, Waterloo, Ontario, N2L 2Y5, Canada}

\affiliation[d]{Computation-Based Science and Technology Research Center, The Cyprus Institute, 20 Kavafi Street, 2121 Nicosia, Cyprus}

\affiliation[e]{Institute of Theoretical Physics and IQST, Universität Ulm, Albert-Einstein-Allee 11, D-89069 Ulm, Germany}

\affiliation[f]{Deutsches Elektronen-Synchrotron DESY, Platanenallee 6, 15738 Zeuthen, Germany}

\emailAdd{angus.kan@uwaterloo.ca}
\emailAdd{lfuncke@mit.edu}
\emailAdd{s.kuehn@cyi.ac.cy}
\emailAdd{luca.dellantonio@uwaterloo.ca}
\emailAdd{jinglei.zhang@uwaterloo.ca}
\emailAdd{jfhaase@uwaterloo.ca}
\emailAdd{christine.muschik@uwaterloo.ca}
\emailAdd{karl.jansen@desy.de}

\abstract{
Quantum and tensor network simulations have emerged as prominent sign-problem free approaches to lattice gauge theories. Unlike conventional Markov chain Monte Carlo methods, they are based on the Hamiltonian formulation.
In this talk, we fill a gap in the literature and present the first derivation of the Hamiltonian 3+1D $\theta$-term---which is an important sign-problem afflicted term---for Abelian and non-Abelian lattice gauge theories.
Furthermore, we perform exact diagonalization for a 3+1D U(1) lattice gauge theory including the $\theta$-term on a unit periodic cube. Our numerical results reveal a novel phase transition at fixed values of $\theta$ in the strong-coupling regime.
The transition is evidenced by an avoided level crossing in the ground state energy, as well as sudden changes in the plaquette expectation value, the electric energy density, and the topological charge density. 
Extensions of our work to larger lattices can be readily performed using state-of-the-art tensor network simulations.
Moreover, our work provides a concrete starting point for an eventual quantum simulation of the $\theta$-dependent phase structure and dynamics of lattice gauge theories in 3+1D.
This talk is mainly based on~\cite{kan2021investigating}. We expand beyond~\cite{kan2021investigating} by including a derivation of the (non-)Abelian fixed-length Higgs term in the Hamiltonian formulation for future studies of (non-)Abelian-Higgs models with a $\theta$-term.\\

Preprint number: MIT-CTP/5350}

\FullConference{%
 The 38th International Symposium on Lattice Field Theory, LATTICE2021
  26th-30th July, 2021
  Zoom/Gather@Massachusetts Institute of Technology
}

\begin{document}
\maketitle

\section{Introduction}
\label{sec:intro}

Lattice gauge theories (LGTs) with topological $\theta$-terms are an important class of sign-problem afflicted problems~\cite{Troyer:2004ge,Gattringer:2016kco} due to their relevance to the strong CP problem. They are largely inaccessible to Markov chain Monte Carlo (MCMC) simulations but can in principle be simulated using tensor networks (TNs) and quantum simulations, performed on classical and quantum processors, respectively. Indeed, 1+1D LGTs with topological $\theta$-terms have been simulated using TNs~\cite{PhysRevD.66.013002,PhysRevD.95.094509,Funcke:2019zna,kuramashi2020tensor,higgs1d,Funcke:2021glr} and a digital quantum computer~\cite{Kharzeev2020real}. Both approaches can in principle be extended to higher dimensions.
TNs have successfully simulated U(1) LGTs in 2+1D~\cite{Felser2019} and 3+1D~\cite{Magnifico:2020bqt}. 
There have been a tremendous progress in quantum simulations of LGTs beyond 1+1D, resulting in a wealth of literature. For a comprehensive overview of LGT quantum simulations, we recommend Refs.~\cite{banuls2020simulating,klco2021standard}, wherein various approaches such as variational, analog, and digital simulations are discussed.
Very recently, efficient quantum algorithms for simulating Abelian and non-Abelian LGTs in any dimensions on universal quantum computers have been developed~\cite{kan2021lattice,tong2021provably}.
Given the rapid developments of sign-problem free algorithms for LGTs in higher dimensions, it is therefore of timely importance to explore lattice $\theta$-terms beyond 1+1D. Unlike conventional MCMC LGT simulations, quantum simulations and many TN methods rely on the Hamiltonian formulation. This is the motivation of our recent work~\cite{kan2021investigating}, where we provided the first derivation of the 3+1D $\theta$-term in the Hamiltonian formulation of LGTs.

In this contribution, we mainly summarize our findings in~\cite{kan2021investigating}. We also expand beyond~\cite{kan2021investigating} by including a derivation of the (non-)Abelian fixed-length Higgs term~\cite{fradkin1979phase} in the Hamiltonian formulation in arbitrary dimensions, see App.~\ref{sec:higgs}. The latter enables extensions of Refs.~\cite{gattringer2015,higgs1d}, where 1+1D Abelian-Higgs models with a $\theta$-term were simulated, to (non-)Abelian-Higgs models in higher dimensions.

We begin by analytically deriving the 3+1D topological $\theta$-term in the Hamiltonian formulation of LGTs using the transfer-matrix method~\cite{creutz1977gauge}. The starting point of our derivation is the Lagrangian definition in Refs.~\cite{Peskin1978,DiVecchia:1981aev}. We then numerically solve a 3+1D compact U(1) pure gauge theory with our $\theta$-term on a single cube using exact diagonalization. We find evidences for a phase transition at constant values of $\theta$, for inverse couplings of $\beta \equiv 1/g^2 \lesssim 0.75$. The evidences include an avoided level-crossing in the ground state energy, and abrupt changes in the plaquette expectation value, the electric energy density, and the topological charge density. 
Our results are in line with previous analytical studies~\cite{Cardy:1981qy,cardy1982duality,honda2020topological} of the $\theta$-dependent phase diagram of a 3+1D pure compact U($1$) LGT, where it was predicted that a phase transition appears at $\theta=\pi$ and small $\beta$ in the confining phase, which disappears at large $\beta$ in the Coulomb phase. Our results also qualitatively match with prior numerical studies in 1+1D~\cite{gattringer2015,Funcke:2019zna,kuramashi2020tensor,higgs1d}. In light of the recent algorithmic developments of TNs~\cite{Magnifico:2020bqt} and quantum simulations~\cite{kan2021lattice,tong2021provably}, our work provides a concrete starting point for a detailed non-perturbative mapping of the $\theta$-dependent phase diagram in 3+1D. 

\section{Derivation of 3+1D \texorpdfstring{\boldmath{$\theta$}}{}-Term in Hamiltonian Formulation}
\label{sec:top}

In this section, we briefly outline our derivation of a 3+1D topological $\theta$-term in the Hamiltonian formulation of LGTs, using the transfer-matrix method~\cite{creutz1977gauge}. We refer the reader to~\cite{kan2021investigating} for details. For clarity, we hereafter disregard the Einstein notation and explicitly display all sums. Further, we distinguish the Lagrangian variables from operators in the Hamiltonian formulation by expressing the latter with a hat ($~\hat{}~$) symbol.

We begin with the Euclidean lattice action
$S=S_W+i\theta \sum_{\vec{n}} Q_{\vec{n}}$, where $S_W$ is Wilson's action, $Q_{\vec{n}}$ is the topological charge, and $\theta$ is the vacuum angle~\cite{bhanot1984lattice}. We employ Peskin's original lattice definition of the topological charge in the Lagrangian formulation~\cite{Peskin1978,DiVecchia:1981aev},
\begin{align}
Q_{\vec{n}} &= -\frac{1}{32\pi^2} \sum_{\mu,\nu,\rho,\sigma}\varepsilon_{\mu\nu\rho\sigma}\,\textrm{Tr}\, \left[ U_{\vec{n},\mu\nu} U_{\vec{n},\rho\sigma}\right],
\label{eq:LTopCharge1}
\end{align}
where $\varepsilon_{\mu\nu\rho\sigma}$ is the 4D Levi-Civita symbol and $U_{\vec{n},\mu\nu}\equiv U_{\vec{n},\mu}U_{\vec{n}+\hat{\mu},\nu}U_{\vec{n}+\hat{\nu},\mu}^\dagger U_{\vec{n},\nu}^\dagger$ is the standard plaquette variable formed by link variables~\cite{wilson1974confinement}. 

To facilitate the use of the transfer-matrix method, we single out the temporal components of $Q_{\vec{n}}$ and rewrite the Euclidean lattice action as
\begin{align}
    S &= S_W - \frac{i\theta}{16\pi^2}\sum_{\vec{n},i,j,k}\varepsilon_{ijk}\text{Tr}\left[\left(U_{\vec{n},0i} - U_{\vec{n},0i}^\dag\right)U_{\vec{n},jk}\right],
    \label{eq:Smod}
\end{align}
where $\varepsilon_{ijk}$ is the 3D Levi-Civita symbol.

In the temporal gauge with $U_{\vec{n},0} = 1$, the transfer matrix can then be written as~\cite{kan2021investigating}
\begin{align}
    \hat{T}
    &= \prod_{\vec{n},i}\int \left(\prod_{b} dx_{\vec{n},i}^b\right) e^{i\sum_b x_{\vec{n},i}^b \hat{E}^{b}_{\vec{n},i}} e^{\frac{a}{2g^2 a_0} \text{Tr}\left[2\cos(\sum_b x_{\vec{n},i}^b\lambda^b)\right]}\nonumber\\
    & \quad \times e^{\frac{i\theta}{16\pi^2}\sum_{j,k}\varepsilon_{ijk}\text{Tr}\left[2i\sin\left(\sum_b x_{\vec{n},i}^b \lambda^b\right)\hat{U}_{\vec{n},jk}\right]} e^{\frac{a_0}{2g^2 a} \sum_{\vec{n},j,k}\text{Tr}\left[\hat{U}_{\vec{n},jk} + \hat{U}_{\vec{n},jk}^\dag\right] },
    \label{eq:Telem}
\end{align}
where $a$ is the spatial lattice spacing, $a_0$ is the temporal lattice spacing, $x_{\vec{n},i}^b \in \mathbbm{R}$ are the group parameters, and $\lambda^b$ are the normalized group generators for the fundamental representation that satisfy $\text{Tr}[\lambda^a\lambda^b]=\delta_{ab}$. Moreover, we have introduced the electric field operators $\hat{E}^{b}_{\vec{n},i}$ that are conjugate to the link operators $\hat{U}_{\vec{n},i}$. They satisfy the commutation relations
\begin{equation}
    \left[\hat{E}^{a}_{\vec{n},i},\hat{E}^{b}_{\vec{n},i}\right] = i\sum_c f^{abc}\hat{E}^{c}_{\vec{n},i},
    \left[\hat{E}^{a}_{\vec{n},i},\hat{U}_{\vec{n},i}\right] = -\lambda^a \hat{U}_{\vec{n},i}\label{eq:link_comm},
\end{equation}
where $f^{abc}$ are the structure constants of the gauge group.

In the temporal continuum limit, as $a_0 \rightarrow 0$, the integral in Eq.~\eqref{eq:Telem} is dominated by the maximum of the cosine term. Thus, we expand the cosine and sine terms in Eq.~\eqref{eq:Telem} around $x_{\vec{n},i}^b = 0$, and obtain a Gaussian integral that evaluates to
\begin{equation}
    \hat{T} \propto e^{-a_0\set*{\hat{H}_{\rm KS}-\frac{ig^2 \theta}{8\pi^2 a}\sum_{\vec{n},i,j,k,b}\varepsilon_{ijk}\text{Tr}\left[\hat{E}_{\vec{n},i}^b \lambda^b\hat{U}_{\vec{n},jk}\right] }},
    \label{eq:TransferResult}
\end{equation}
where $\hat{H}_{\rm KS}$ is the Kogut-Susskind pure gauge Hamiltonian~\cite{kogut1975hamiltonian},
\begin{equation}
    \hat{H}_{\rm KS} = \frac{g^2}{2a} \sum_{\vec{n},i,b} \hat{E}^b_{\vec{n},i}\hat{E}^b_{\vec{n},i} - \frac{1}{2g^2 a} \sum_{\vec{n},j,k}\text{Tr}\left[\hat{U}_{\vec{n},jk} + \hat{U}_{\vec{n},jk}^\dag\right].
    \label{eq:KogutSusskind}
\end{equation}
Since the transfer matrix is defined as $\hat{T} \equiv e^{-a_0 \hat{H}}$, where $\hat{H}$ is the Hamiltonian, we can directly read off the topological $\theta$-term in the Hamiltonian formulation,
\begin{align}
    \theta\hat{Q}=-\frac{ig^2 \theta}{8\pi^2 a}\sum_{\vec{n},i,j,k,b}\varepsilon_{ijk}\text{Tr}\left[\hat{E}_{\vec{n},i}^b \lambda^b\hat{U}_{\vec{n},jk}\right] = -\frac{ig^2 \theta}{4\pi^2 a}\sum_{\vec{n},b} \sum_{(i,j,k)\in \text{even}}\text{Tr}\left[\hat{E}_{\vec{n},i}^b \lambda^b\left(\hat{U}_{\vec{n},jk}-\hat{U}_{\vec{n},jk}^\dag\right)\right],
    \label{eq:top1}
\end{align}
where $\varepsilon_{ijk}$ is the 3D Levi-Civita symbol, $(i,j,k)$ is summed over the set of even permutations, and we have used $\hat{U}_{\vec{n},jk} = \hat{U}_{\vec{n},kj}^\dag$. 
In order to better approximate the electric field at each site, we replace the outgoing electric field $\hat{E}_{\vec{n},i}^b$ with an average of the incoming and outgoing electric fields. As such, we obtain the following improved definition of the topological $\theta$-term:
\begin{equation}
    \theta \hat{Q} = -\frac{ig^2 \theta}{8\pi^2 a}\sum_{\vec{n},b} \sum_{(i,j,k)\in \text{even}}
    \text{Tr}\left[\left(\hat{E}_{\vec{n}-\hat{i},i}^b+\hat{E}_{\vec{n},i}^b\right) \lambda^b\left(\hat{U}_{\vec{n},jk}-\hat{U}_{\vec{n},jk}^\dag\right)\right].
    \label{eq:TopTermSymm}
\end{equation}

\section{Model and Methods}
\label{sec:model}

In this section, we apply the expression in Eq.~\eqref{eq:TopTermSymm} to a 3+1D U(1) lattice gauge theory. Using exact diagonalization, we numerically investigate the theory at non-vanishing values of $\theta$ on a single cube with periodic boundary conditions. In our simulations, we consider the Hamiltonian
\begin{align}
    \hat{H} &= \hat{H}_E + \hat{H}_B + \tilde{\theta} \hat{Q}, \quad \hat{H}_E = \frac{1}{2\beta}\sum_{\vec{n}}\sum_{j=1}^{3} \hat{E}_{\vec{n},j}^2, \quad 
    \hat{H}_B = -\frac{\beta}{2} \sum_{\vec{n}}\sum_{j,k=1;k>j}^{3} (\hat{U}_{\vec{n},jk} + \hat{U}_{\vec{n},jk}^\dag ), \nonumber \\
    \tilde{\theta} \hat{Q} &= -i \frac{\tilde{\theta}}{\beta} \sum_{\vec{n}}\sum_{(i,j,k)\in \text{even}} (\hat{E}_{\vec{n}-\hat{i},i}+\hat{E}_{\vec{n},i})(\hat{U}_{\vec{n},jk} - \hat{U}_{\vec{n},jk}^\dag), \label{eq:U1_Q}
\end{align}
where we have set $a=1$, $\beta = 1/g^2$, $\tilde{\theta} = \theta/8\pi^2$, and $(i,j,k)$ is summed over the set of even permutations. Moreover, the link operators satisfy $[\hat{E}_{\vec{n},j}, \hat{U}_{\vec{n}',j'}] = \delta_{\vec{n},\vec{n}'}\delta_{j,j'} \hat{U}_{\vec{n},j}$.
We use eigenstates of the electric field operators
$\hat{E}_{\vec{n},j}\ket{{E}_{\vec{n},j}}= {E}_{\vec{n},j}\ket{{E}_{\vec{n},j}}, \: {E}_{\vec{n},j} \in \mathbbm{Z}$
as a basis for the gauge fields.
In order to simulate the infinite-dimensional gauge-field operators on a computer with finite resources, we truncate these operators at a cutoff $s$ such that they become
\begin{equation}
    \hat{E}_{\vec{n},j} = \sum_{{E}_{\vec{n},j}=-s}^{s}{E}_{\vec{n},j}\ketbra{{E}_{\vec{n},j}}{{E}_{\vec{n},j}}, \quad 
    \hat{U}_{\vec{n},j} = \sum_{{E}_{\vec{n},j}=-s+1}^{s}\ketbra{{E}_{\vec{n},j}-1}{{E}_{\vec{n},j}}.
\end{equation}
Hereafter, we use the minimal symmetric truncation and set $s=1$. 

The Hilbert space that is spanned by the electric eigenstates contains many unphysical states that violate Gauss' law, i.e.,
\begin{equation}
    \hat{G}_{\vec{n}} = \sum_{i = 1}^{3}(\hat{E}_{\vec{n},i}-\hat{E}_{\vec{n}-\hat{i},i}), \:\hat{G}_{\vec{n}}\ket{\Psi} = 0, \: \forall \vec{n}.
\end{equation}
Thus, the spectrum of the Hamiltonian will be contaminated by these unphysical states. Following~\cite{PhysRevD.102.094515,haase2021resource}, we overcome this issue by deriving and simulating an effective Hamiltonian, where Gauss' law is incorporated as a set of constraints over the gauge-field operators. The resulting effective Hamiltonian not only is supported by a physical, gauge-invariant subspace, but also requires less memory to simulate. For a 3D periodic cubic lattice with $N=L^3$ sites, the number of basis states is reduced from $(2s+1)^{3L^3}$ to $(2s+1)^{2L^3-1}$.

\section{Numerical Results}
\label{sec:results}

\begin{figure*}[htp!]
\centering
\begin{tabular}{cc}
  \includegraphics[width=0.49\linewidth]{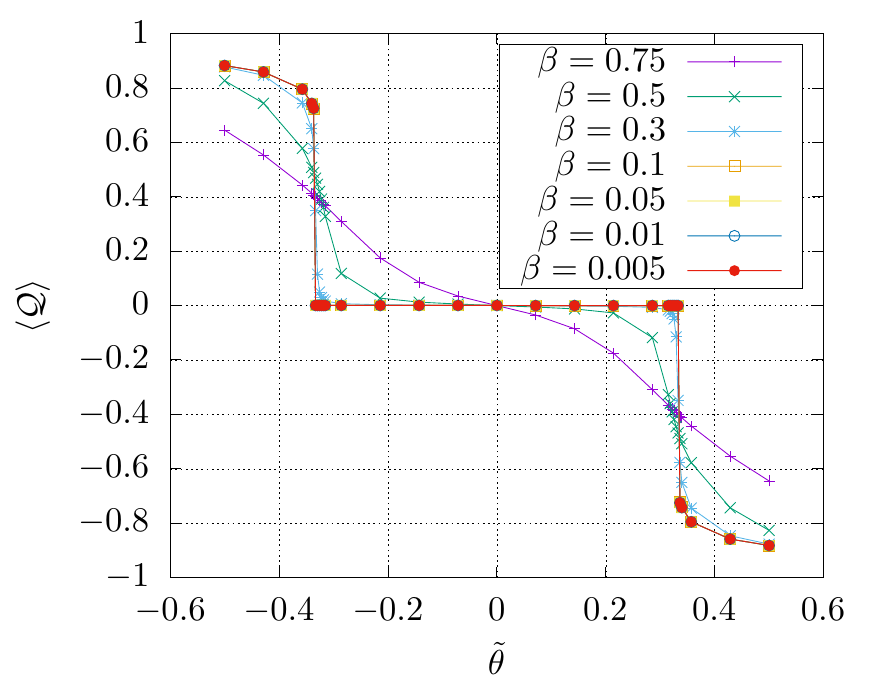} & \includegraphics[width=0.49\linewidth]{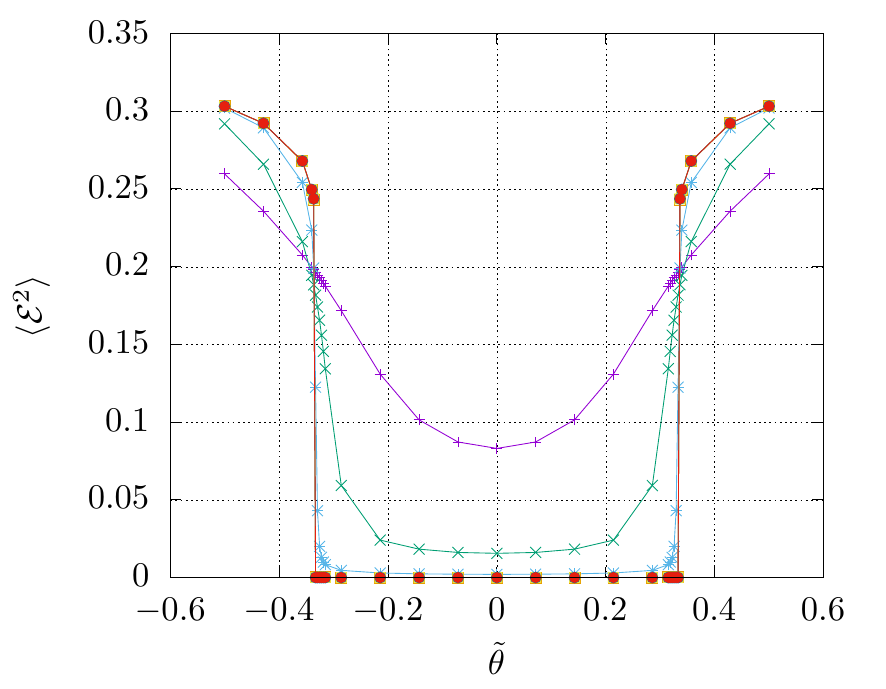}    \\
 (a) & (b) \\[6pt]
  \includegraphics[width=0.49\linewidth]{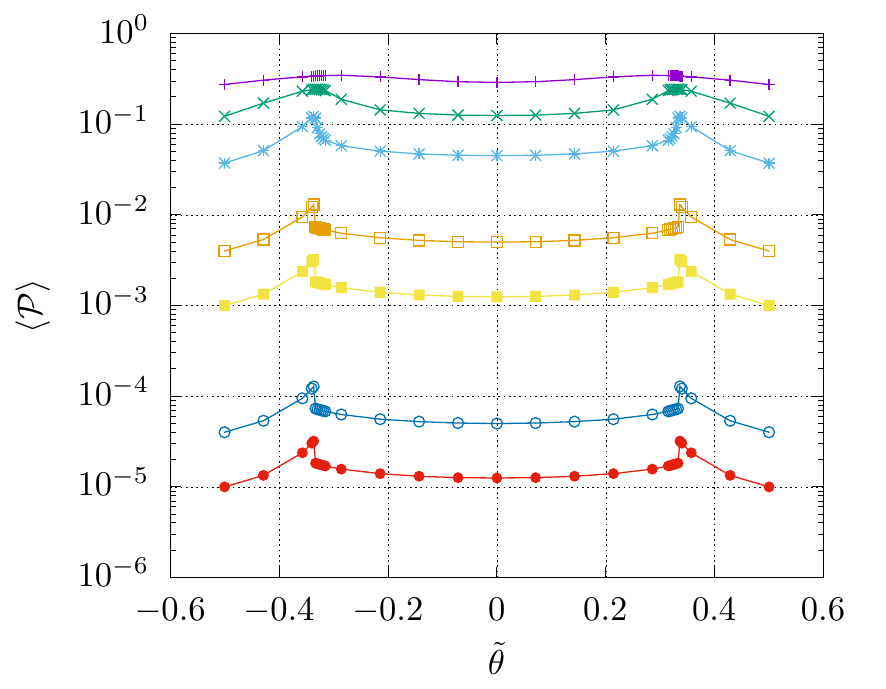}& \includegraphics[width=0.49\linewidth]{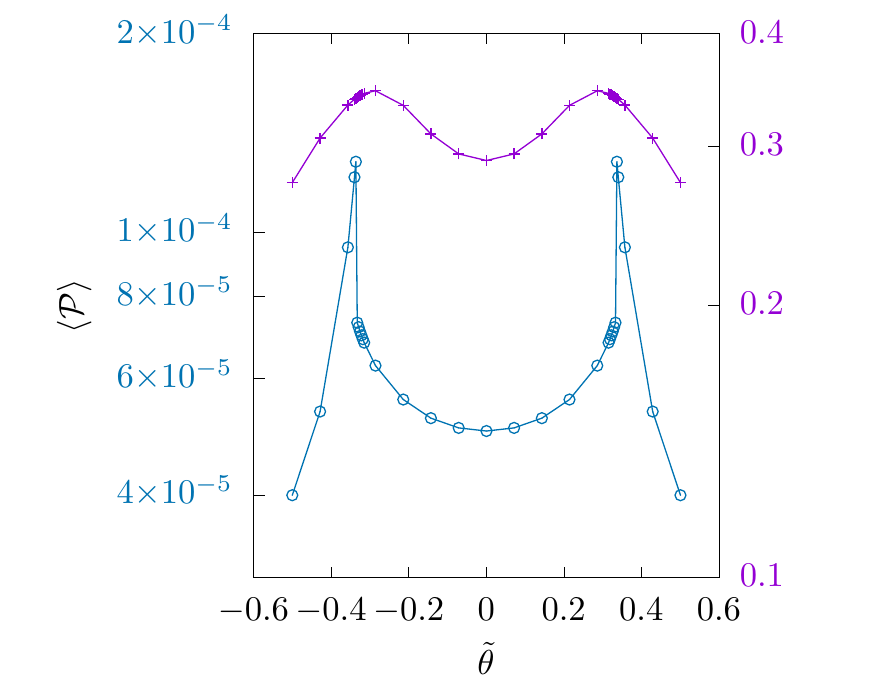}  \\ 
 (c) & (d) \\[6pt]
\end{tabular}
\caption{(a) Bare topological charge density, (b) bare electric energy density, and (c) plaquette expectation value as a function of $\tilde{\theta}$ for $\beta \leq 0.75$. (d) The plaquette expectation values for $\beta = 0.01$ (left $y$-axis) and $0.75$ (right $y$-axis) are shown in greater detail to emphasize the change in the behavior as $\beta$ increases. In (a) and (b), note that the lines for $\beta=0.1,0.05,0.01$ are covered by the red line. This figure is taken from~\cite{kan2021investigating}.}
\label{fig:observables}
\end{figure*}

In this section, we report our numerical findings in~\cite{kan2021investigating}.
We use exact diagonalization to compute the low-lying spectrum of the Hamiltonian in Eq.~\eqref{eq:U1_Q} on a single periodic cube. We focus on the $\theta$-dependence of the energy spectrum and of the ground-state expectation value of various observables. In particular, we study the plaquette expectation value
\begin{equation}
    \expval{\mathcal{P}} = -\frac{1}{V\beta} \langle \Psi_0 |\hat{H}_B| \Psi_0\rangle,
    \label{eq:plaq}
\end{equation}
where $V$ is the number of plaquettes in the lattice and $\ket{\Psi_0}$ is the ground state, the bare topological charge density
\begin{equation}
    \expval{\mathcal{Q}} = -\frac{\beta}{V} \langle \Psi_0 |\hat{Q}| \Psi_0\rangle,
\end{equation}
and the bare electric energy density
\begin{equation}
    \expval{\mathcal{E}^2} = \frac{\beta}{V} \langle \Psi_0 |\hat{H}_E| \Psi_0\rangle.
\end{equation}

We summarize our results for the topological charge density, the electric energy density, and the plaquette expectation value in Fig.~\ref{fig:observables}. We simulated a wide range of couplings $\beta \in [0.005,0.75]$, and for each value of $\beta$, a range of values $\tilde{\theta} \in [-0.5,0.5]$. Both the topological charge density and the electric energy density exhibit discontinuities at $|\tilde{\theta}| \approx 0.333$ when $\beta \leq 0.3$, as shown in Figs.~\ref{fig:observables}(a) and \ref{fig:observables}(b). The plaquette expectation value displays spikes at these points, as depicted in Figs.~\ref{fig:observables}(c) and \ref{fig:observables}(d). These abrupt changes point towards a phase transition occurring at $|\tilde{\theta}| \approx 0.333$. Also shown in Fig.~\ref{fig:observables}, these distinct features vanish as $\beta$ increases beyond $0.3$, suggesting the disappearance of the phase transition in this regime.

Our observed phase transition most likely corresponds to the theoretically predicted phase transition at $\theta=\pi$~\cite{cardy1982duality,honda2020topological}. The main reason is that this transition, to our knowledge, is the only analytically predicted constant-$\theta$ transition in a compact U(1) gauge theory. Furthermore, in agreement with our observed transition, the analytical studies~\cite{cardy1982duality,honda2020topological} predict the occurrence of this transition for small $\beta$, which vanishes at large $\beta$ as one approaches the Coulomb phase. 
However, since we only simulate a single cube, which is the minimal volume in 3+1D, harsh finite-volume effects are expected. 
In particular, we attribute the shift in our observed transition point from $\theta=\pi$ and the lack of the predicted $2\pi$-periodicity in $\theta$~\cite{cardy1982duality,honda2020topological} to finite-volume effects.
For evidences, we turn to related numerical studies of 1+1D U(1) models with a $\theta$-term~\cite{gattringer2015,Funcke:2019zna,kuramashi2020tensor,higgs1d}, since, to our knowledge, similar studies to ours in 3+1D do not exist. A lack of periodicity in $\theta$ and noticeable shifts of the $\theta=\pi$ transition point to larger $\theta$-values were reported for simulations of lattices with only a few sites, i.e., $\lesssim 10$~\cite{gattringer2015,higgs1d}. 
In~\cite{kuramashi2020tensor}, a discontinuity in the topological charge density emerged only when the lattice size reached $64$ sites, and was not observed in lattices with $\leq 32$ sites.
Only in simulations of larger lattices with $\gtrsim 100$ sites, both the $2\pi$-periodicity and the $\theta=\pi$ transition point are fully restored~\cite{gattringer2015,Funcke:2019zna,kuramashi2020tensor,higgs1d}. As such, in 3+1D, we expect to recover both the periodicity and the correct transition point in larger lattices as well. In the future, this could be verified by including our $\theta$-term in the TN simulations for 3+1D U(1) LGTs~\cite{Magnifico:2020bqt}, as well as eventual quantum simulations of LGTs. 
We note in passing that the truncation of electric fields could have plausibly played a role in the lack of periodicity and the shift in transition point. However, such truncation effects were shown to be neglibigle in~\cite{kan2021investigating}.
Additionally, we note that in our simulations, the observables obey the following symmetries: $\expval{\mathcal{P}(\theta)} = \expval{\mathcal{P}(-\theta)}$, $\expval{\mathcal{E}^2(\theta)} = \expval{\mathcal{E}^2(-\theta)}$, and $\expval{\mathcal{Q}(\theta)} = -\expval{\mathcal{Q}(-\theta)}$. Similar symmetries were observed in Refs.~\cite{gattringer2015,Funcke:2019zna,kuramashi2020tensor}. 
We remark that these symmetries can be directly inferred from the fact that the CP transformation $U_{\vec{n},i}\rightarrow U_{\vec{n},i}^\dag$ flips the sign of the topological charge but not that of the plaquette expectation and electric energy density.

\begin{figure}[htp!]
\centering
    \includegraphics[width=0.5\linewidth]{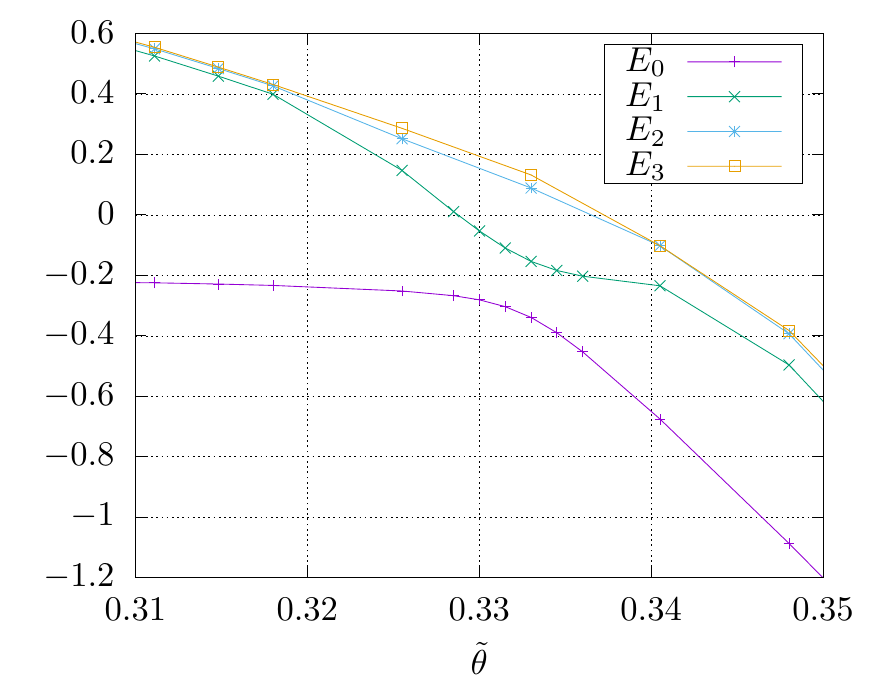}
    \caption{Low-lying spectrum as a function of $\tilde{\theta}$ for $\beta = 0.3$. The ground state and the first excited state exhibit an avoided level-crossing at $\tilde{\theta} \approx 0.333$. This figure is taken from~\cite{kan2021investigating}.}
    \label{fig:crossing}
\end{figure}

Next, we investigate the order of the quantum phase transition, which is unknown from analytical predictions~\cite{Cardy:1981qy,cardy1982duality}. We focus on the low-lying energy spectrum of the Hamiltonian at $\beta=0.3$ and near the transition point $\tilde{\theta} \approx 0.333$. Our results for the first four energy levels, shown in Fig.~\ref{fig:crossing}, reveal an avoided level-crossing between the ground state and the first excited state at the transition point. These findings rule out a first-order transition, where one would expect a level crossing. Instead, the spectrum suggests that a second or higher-order phase transition is causing the sudden changes in the observables in Fig.~\ref{fig:observables}.

\section{Conclusion and Outlook}
\label{sec:conclusion}

In this talk, we presented the transfer-matrix derivation of the Hamiltonian lattice $\theta$-term for 3+1D Abelian and non-Abelian LGTs, based on our recent work~\cite{kan2021investigating}. We then discussed the numerical results obtained in~\cite{kan2021investigating}, where a 3+1D U(1) LGT with our $\theta$-term on a single periodic cube was diagonalized. Our numerical results indicate a phase transition at constant values of $\theta$ and small values of $\beta$, which eventually vanishes as $\beta$ increases towards the Coulomb phase. These findings are in accordance with the analytically predicted phase transition of Refs.~\cite{cardy1982duality,honda2020topological}. Finally, our simulations reveal an avoided level-crossing, implying that this transition is not of first order, which was previously unknown to Refs.~\cite{cardy1982duality,honda2020topological}.

Using the TN methods developed for 3+1D LGTs in Ref.~\cite{Magnifico:2020bqt}, our numerical results can be readily cross-checked and extended to larger lattices. Moreover, the 1+1D Abelian-Higgs study may be extended to 3+1D by including the 3+1D U(1) Higgs term in~\cite{gonzalez2017quantum}. To further enable non-Abelian-Higgs studies, we straightforwardly generalize the derivation in~\cite{gonzalez2017quantum} to the non-Abelian case in App.~\ref{sec:higgs}, again using the transfer-matrix method~\cite{creutz1977gauge}.

In the light of the recently developed quantum algorithms for LGTs~\cite{kan2021lattice,tong2021provably}, we envision future fault-tolerant quantum simulations of 3+1D LGTs including our $\theta$-term.
To build towards this vision, one can create efficient quantum circuits to simulate our $\theta$-term following Ref.~\cite{kan2021lattice}. One begins by expanding the electric field and link operators in the group representation basis~\cite{byrnes2006simulating,zohar2015formulation}, as we have done in this work for the U(1) case. The electric field operators in this basis for the SU(2) and SU(3) cases were shown in Refs.~\cite{robson1982gauge} and~\cite{mukunda1965tensor}, respectively. The SU(2) and SU(3) link operators in the fundamental representation in this basis can be found in Ref.~\cite{byrnes2006simulating}. As such, the circuit synthesis techniques in Ref.~\cite{kan2021lattice} can be applied to construct gate-by-gate quantum circuits to simulate our $\theta$-term.

\section*{Acknowledgements}

L.F.\ thanks A. Avkhadiev, M. Creutz and D. Kharzeev for discussions, and is partially supported by the U.S.\ Department of Energy, Office
of Science, National Quantum Information Science Research Centers,
Co-design Center for Quantum Advantage (C$^2$QA) under contract number
DE-SC0012704, by the DOE QuantiSED Consortium under subcontract number
675352, by the National Science Foundation under Cooperative Agreement
PHY-2019786 (The NSF AI Institute for Artificial Intelligence and
Fundamental Interactions, http://iaifi.org/), and by the U.S.\
Department of Energy, Office of Science, Office of Nuclear Physics under grant contract numbers DE-SC0011090 and DE-SC0021006.
S.K.\ acknowledges financial support from the Cyprus Research and Innovation Foundation under project “Future-proofing Scientific Applications for the Supercomputers of Tomorrow (FAST)”, contract no.\ COMPLEMENTARY/0916/0048. 
J.F.H.\ acknowledges the Alexander von Humboldt Foundation for a Feodor Lynen Fellowship.
C.A.M.\ acknowledges the Alfred P. Sloan foundation for a Sloan Research Fellowship.
This work is supported in part by Transformative Quantum Technologies Program (CFREF), NSERC, New frontiers in Research Fund, Compute Canada, and European Union’s Horizon 2020 research and innovation programme under the Grant Agreement No.\ 731473 (FWF QuantERA via QTFLAG I03769). Research at Perimeter Institute is supported in part by the Government of Canada through the Department of Innovation, Science and Industry Canada and by the Province of Ontario through the Ministry of Colleges and Universities.

\appendix

\section{Derivation of the fixed-length Higgs term}
\label{sec:higgs}

In this section, we derive the fixed-length Higgs term for (non-)Abelian LGTs in the Hamiltonian formulation in arbitrary dimensions, using the transfer-matrix method~\cite{creutz1977gauge}.
We start with the action for a fixed-length Higgs field in the fundamental representation~\cite{fradkin1979phase},
\begin{equation}
    S = -\frac{\kappa a^{d-1}}{2}\sum_{\vec{n},\mu} \text{Tr}\left[\Phi_{\vec{n}} U_{\vec{n},\mu} \Phi_{\vec{n}+\hat{\mu}}^\dag + h.c.\right],
    \label{eq:higgs}
\end{equation}
where $\kappa$ is the coupling, $d$ is the spatial dimension, and the Higgs variable ${\Phi}_{\vec{n}}$, similar to $U_{\vec{n},\mu}$, is a gauge group matrix in the fundamental representation. 
Note the action has the same structure, i.e., a sum of products of gauge group matrices, as the standard Wilson's action~\cite{wilson1974confinement}. 
Particularly in the temporal gauge, where $U_{\vec{n},0} = 1$, the temporal parts of the two actions become mathematically identical, up to differences in the prefactors, i.e., $\sum_{\vec{n}}\text{Tr}[\Phi_{\vec{n}}^\prime \Phi_{\vec{n}}^\dag + h.c.]$ versus $\sum_{\vec{n},i}\text{Tr}[U_{\vec{n},i}^\prime U_{\vec{n},i}^\dag + h.c.]$, where the variables with and without the prime $({ }^\prime)$ are from consecutive time slices.
This motivates us to directly follow the derivation of the Kogut-Susskind Hamiltonian from Wilson's action in Ref.~\cite{creutz1977gauge}.
We define a pair of conjugate site operators, $\hat{\Pi}_{\vec{n}}^a$ and $\hat{\Phi}_{\vec{n}}$, which satisfy the relations
\begin{equation}
    \left[\hat{\Pi}_{\vec{n}}^a,\hat{\Pi}_{\vec{n}}^b\right] = i\sum_c f^{abc}\hat{\Pi}_{\vec{n}}^c, \quad 
    [\hat{\Pi}_{\vec{n}}^a,\hat{\Phi}_{\vec{n}}] = -\lambda^{a}\hat{\Phi}_{\vec{n}}\label{eq:site_comm}.
\end{equation}
In terms of the site and link operators, we then obtain the transfer matrix
\begin{equation}
    \hat{T}= \prod_{\vec{n}} \int \left(\prod_{a}dx^a_{\vec{n}}\right) e^{i\sum_b \hat{\Pi}_{\vec{n}}^b x_{\vec{n}}^b}e^{ \frac{\kappa a^d}{2a_0} \text{Tr}\left[2\cos(\sum_b \lambda^b x_{\vec{n}}^b)\right] } e^{\frac{\kappa a_0 a^{d-2}}{2} \sum_{\vec{m},j} \text{Tr}\left[\hat{\Phi}_{\vec{m}} \hat{U}_{\vec{m},j} \hat{\Phi}_{\vec{m}+\hat{j}}^\dag + h.c.\right]}.
\end{equation}
Note that this expression has essentially the same form as the transfer matrix in Eq.~(5.25) in Ref.~\cite{creutz1977gauge}. In both cases, the temporal part of each action gives rise to an exponentiated cosine term.
When $a_0 \rightarrow 0$, the integral is dominated by the region near the maximum of cosine function, where $x_{\vec{n}} \approx 0$. Expanding about $x_{\vec{n}} = 0$ and keeping only the quadratic term, as in Ref.~\cite{creutz1977gauge}, the transfer matrix becomes a Gaussian integral that evaluates to
\begin{equation}
    \hat{T} \propto e^{-a_0 \left\{ \frac{1}{2\kappa a^d} \sum_{\vec{n}}\hat{\Pi}^2_{\vec{n}}-\frac{\kappa a^{d-2}}{2} \sum_{\vec{m},j} \text{Tr}\left[\hat{\Phi}_{\vec{m}} \hat{U}_{\vec{m},j} \hat{\Phi}_{\vec{m}+\hat{j}}^\dag + h.c.\right] \right\} }.
\end{equation}
From this, we can directly read off the Higgs term
\begin{equation}
    \frac{1}{2\kappa a^d} \sum_{\vec{n}}\hat{\Pi}^2_{\vec{n}}-\frac{\kappa a^{d-2}}{2} \sum_{\vec{n},j} \text{Tr}\left[\hat{\Phi}_{\vec{n}} \hat{U}_{\vec{n},j} \hat{\Phi}_{\vec{n}+\hat{j}}^\dag + h.c.\right].
\end{equation}
When applied to the U(1) gauge group, this reduces to the Abelian-Higgs term derived in~\cite{gonzalez2017quantum}.


\bibliographystyle{JHEP}
\bibliography{Papers}
\end{document}